\documentclass{aastex}          %% The manuscript based on AASTeX v5.x
\usepackage{spr-astr-addons}    %% mimicing ASTR journal style
\begin{document}
%% Article title
%
\title{Inclusion of sdBs in evolutionary population synthesis for binary stellar populations and the application: the determinations of photo-$z$ and galaxy morphology}

%% Running heads
\shorttitle{Inclusion of sdBs in EPS for BSPs and the application: the determinations of photo-$z$ and galaxy morphology}
\shortauthors{Zhang et al.}

%% Author and Affilations
\author{F. Zhang\altaffilmark{1,3,4}}
\and
\author{Z. Han\altaffilmark{1,3}}
\and
\author{L. Li\altaffilmark{1,3}}
\and
\author{J. Guo\altaffilmark{1,2,3}}
\and
\author{Y. Zhang\altaffilmark{1,2,3}}
%\affil{Yunnan observatory}
\email{zhangfh@ynao.ac.cn} %% non-output

%% Alternate Affilations
\altaffiltext{1}{National Astronomical Observatories/Yunnan
Observatory, Chinese Academy of Sciences, Kunming, 650011,
P.R.China.}
\altaffiltext{2}{Graduate University of Chinese Academy
of Sciences, Beijing 100039, China.}
\altaffiltext{3}{Key Laboratory for the Structure and Evolution
of Celestial Objects, Chinese Academy of Sciences.}
\altaffiltext{4}{E-mail:zhang\_fh@hotmail.com or zhangfh@ynao.ac.cn}
%\altaffiltext{3}{}

%% Abstract
\begin{abstract}
Subdwarf B stars (sdBs) can { significantly change} the ultraviolet spectra of populations at age $t\sim$1\,Gyr, and have been even included in { the} evolutionary population synthesis (EPS) models by Han et al. (2007).
In this study we present the spectral energy distributions (SEDs) of binary stellar populations (BSPs) by combining the EPS models of Han et al. (2007) and those of { the} Yunnan group (Zhang et al. 2004, 005), which have included various binary interactions (except sdBs) in EPS models. This set { of} SEDs { is} available { upon} request from the authors.

Using this set { of} SEDs of BSPs we build the spectra of Burst, E, S0-Sd and Irr types of galaxies by using the package of Bruzual \& Charlot { (2003, BC03)}. { Combined} with the photometric data (filters and magnitudes), we obtain the photometric redshifts and morphologies of 1502 galaxies by using the \textit{Hyperz} code of Bolzonella et al. { (2000)}. This sample of galaxies is obtained by removing those objects, { mismatched} with {\sl the} SDSS/DR7 and GALEX/DR4, from the catalogue of Fukugita et al. { (2007)}. By comparison the results with the SDSS spectroscopic redshifts and the morphological index of Fukugita et al. { (2007)}, we find that the photo-$z$s fluctuate with the SDSS spectroscopic redshifts, while the Sa-Sc galaxies in the catalogue of Fukugita et al. { (2007)} are classified earlier as Burst-E galaxies.
\end{abstract}

%% Keywords

%____________________________________________________________________________
%____________________________________________________________________________
\section{Introduction}\label{s:intro}
Binary interactions play an important role in evolutionary population synthesis (EPS) models, { and} their inclusion can affect the overall shape of the spectral energy distribution (SED) of population. In particular, the SED in the ultraviolet passbands is bluer by 2-3 magnitude at $t\sim$ 1\,Gyr if binary interactions are included. The bluer SED in the ultraviolet passbands is caused by { those} binary interactions { which} can create some important classes of objects, such as hot subdwarf B stars (sdBs) for { a} population older than $\sim$1\,Gyr (Han, Podsiadlowski \& Lynas-Gray 2007, hereafter HPL07) and blue stragglers at 0.5$\sim$1.5Gyr.

{ The} Yunnan group (Zhang et al. 2004, 2005) has included various binary interactions in EPS models and { has} presented the SEDs and colours for an extensive set of binary stellar populations (BSPs), { however, it has neglected sdBs}.
Recently, Han et al. (2002, 2003) proposed a binary model for the formation of hot sdBs in binaries and single hot sdBs (three formation channels), and HPL07 presented the SEDs of populations with the binary interactions of sdBs.
In this study we will present the new set { of} SEDs (HPL07+Yunnan models) for BSPs by combining the EPS models of HPL07 with those of Yunnan models.

As an application of this { new} set { of} SEDs of BSPs, firstly, we build the spectra of Burst, E, S0, Sa-Sd{ ,} and Irr types of galaxies{ . Combing them} with the \textit{Galaxy Evolution Explorer} (GALEX) $F_{\rm UV}$, $N_{\rm UV}$, \textit{Sloan Digital Sky Survey} (SDSS) $ugriz$ magnitudes and errors of galaxies, we obtain their photometric redshifts (photo-$z$s) and morphologies by using the \textit{Hyperz} code of Bolzonella, Miralles \& Pello (2000). These galaxies are from the catalogue of Fukugita et al. (2007) and matched with {\sl the} GALEX Data Release four (DR4) and SDSS Data Release seven (DR7).
The outline of the paper is as follows. In Section 2 we present the new set { of} SEDs of BSPs. In Section 3 we apply this set of SEDs to the determinations of photo-$z$ and galaxy morphology.

%____________________________________________________________________________
%____________________________________________________________________________
\section{Inclusion of sdBs in EPS for BSPs}
\begin{figure}[t]
\begin{center}
\includegraphics[height=6.0cm,width=6.0cm,angle=270]{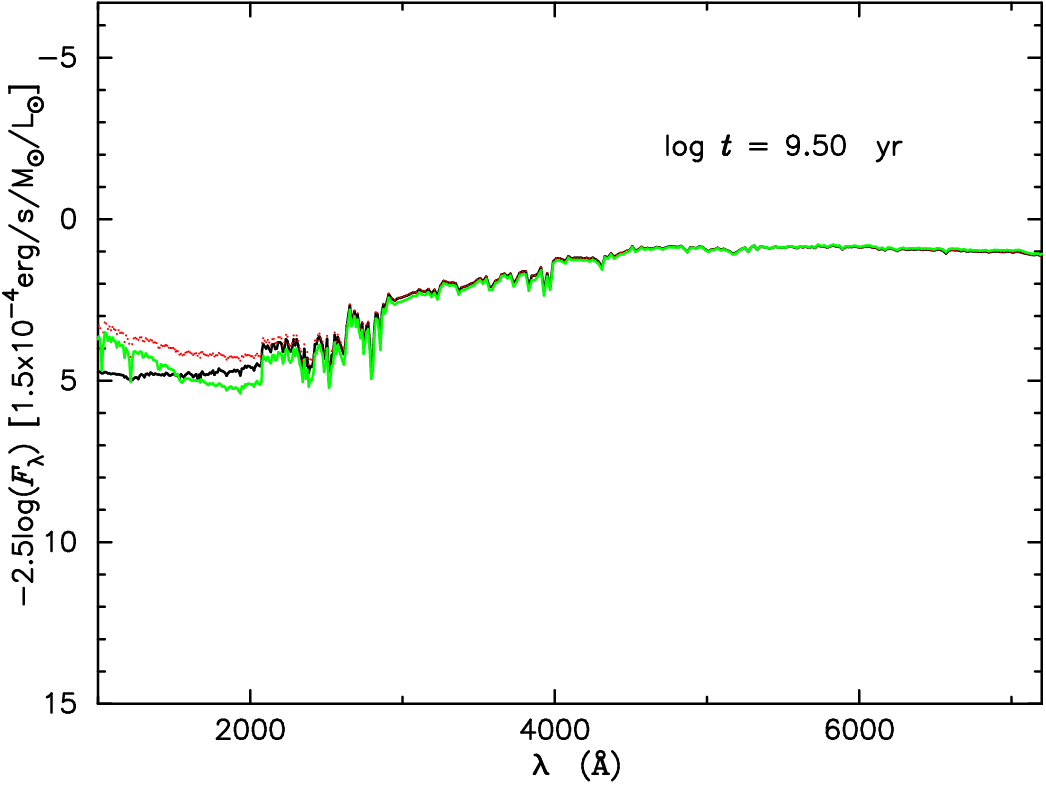}
\caption{The SED of {\sl the} Yunnan (black), HPL07 (green) and Yunnan+HPL07 (red) models for a solar-metallicity populations (normalized to 1$M_\odot$) at age log($t$)= 9.50 yr.}
\end{center}
\end{figure}

\begin{figure}[t]
\begin{center}
\includegraphics[height=6.0cm,width=6.0cm,angle=270]{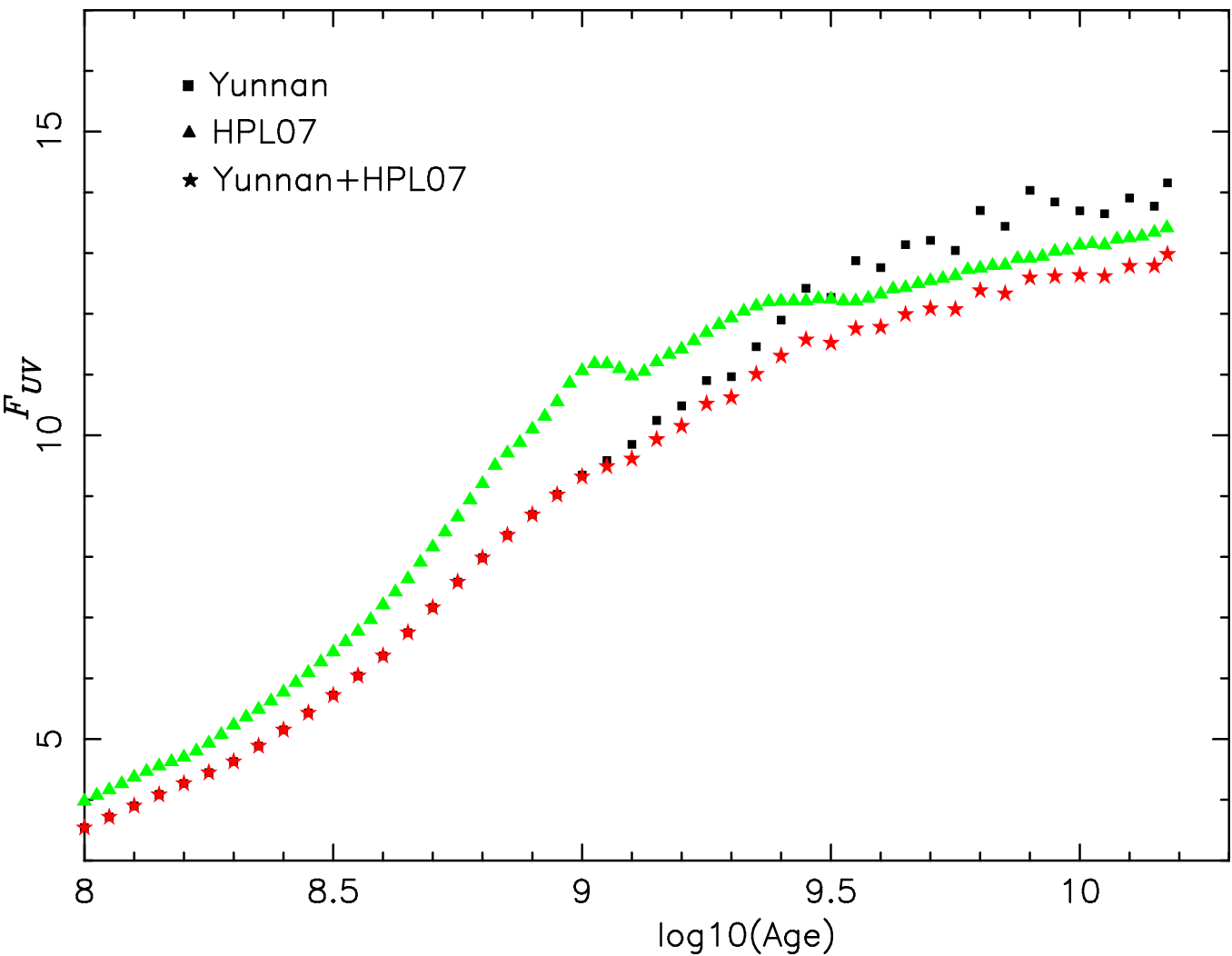}
\caption{Evolution of GALEX $F_{\rm UV}$ magnitude of a solar-metallicity population (1$M_\odot$) for {\sl the} Yunnan (rectangles), HPL07 (triangles) and HPL07+Yunnan (stars) models.}
\end{center}
\end{figure}

In this study we use the standard HPL07 and Yunnan EPS models.
In the standard Yunnan models { are the following assumptions:}
(i) the initial mass-ratio distribution satisfies a uniform { value}, i.e., $n(q)=1$;
(ii) the common envelope (CE) ejection efficiency $\alpha'_{\rm CE} = 3.0$, which is introduced { with} the following criterion { (in the CE evolution model)}: the CE is ejected when the change in orbital energy ($\Delta E_{\rm orb}$), multiplied by $\alpha'_{\rm CE}${ ,} exceeds the change in binding energy of the envelope ($\Delta E_{\rm bind}$); and{ ,}
(iii) the Reimers wind mass-loss coefficient $\eta$ is 0.3.
The ages of BSPs are in the range 5.0 $\le {\rm log}(t{\rm /yr}) \le 10.2$, and the metallicities are in the range -2.3 $\le$ [Fe/H] $\le$ 0.3.

In the standard HPL07 models{ , we have the following:} (i) $n(q)=1$; (ii) the critical mass ratio $q_{\rm c}$ is 1.5, above which the first Roche lobe overflow on the first giant branch or asymptotic giant branch is unstable; and
(iii) $\alpha_{\rm CE}=0.75$ and $\alpha_{\rm th}=0.75$, { assuming that} when $\alpha_{\rm CE} \cdot \Delta E_{\rm orb} + \alpha_{\rm th} \cdot \Delta E_{\rm th}$ (the change in thermal energy of the envelope) exceeds $\Delta E_{\rm bind}${ ,} the CE is ejected. In this { prescription} the thermal energy of the envelope is considered.
The HPL07 models are only for solar-metallicity populations, and the ages of populations are in the range 8.0 $\le {\rm log}(t{\rm /yr}) \le 10.2$.

Combining the EPS models of { the} Yunnan group with those of HPL07, we obtain a new set of SEDs for solar-metallicity BSPs (hereafter Yunnan+HPL07 models). The ages are in the range 5.0 $\le {\rm log}(t{\rm /yr}) \le 10.2$ at a logarithmic age interval of 0.05 or 0.1.

In Figs.1 and 2 we give the SED at age log($t$/yr) = { 9.50} and the evolution of GALEX $F_{\rm UV}$ magnitude of a solar-metallicity population with a mass of 1\,M$_{\odot}$ for Yunnan, HPL07 and  HPL07+Yunnan models. From it we see that the ultraviolet spectra and $F_{\rm UV}$ magnitude of HPL07+Yunnan models are significantly bluer than those of { the} Yunnan and HPL07 models at late ages.

%____________________________________________________________________________
\section{Application to the determination of photo-$z$ and galaxy morphology}\label{s:appli}
\begin{figure}[t]
\begin{center}
\includegraphics[height=6.0cm,width=6.0cm,angle=270]{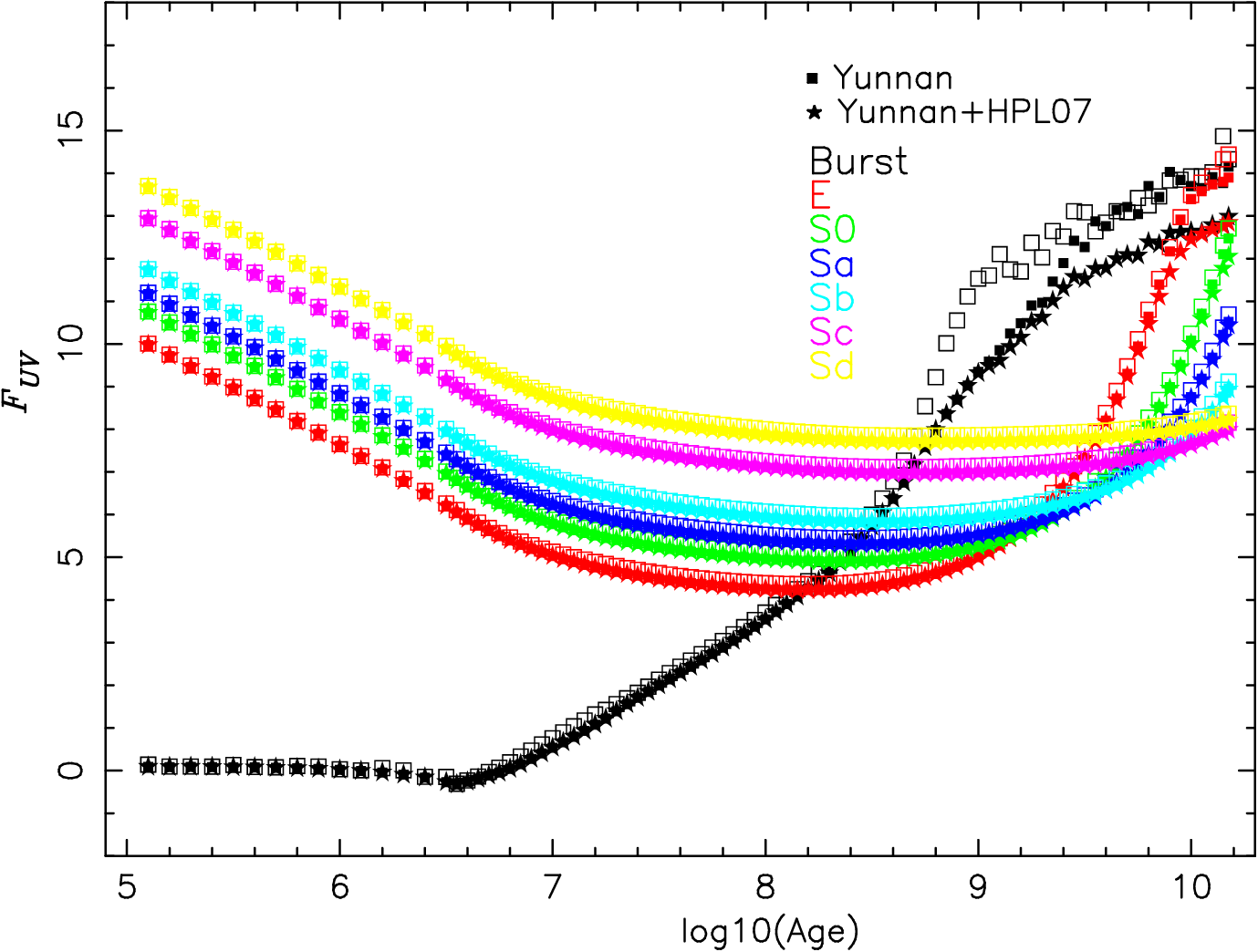}
\caption{Evolution of $F_{\rm UV}$ magnitude of Burst (black), E (red), S0 (green), Sa (blue), Sb (light blue), Sc (carmine) and Sd (yellow) galaxies  (1$M_\odot$) which are built with the Yunnan+HPL07 (stars) and Yunnan (solid rectangles) models. Also shown are the results built with the Yunnan models without any binary interactions (open rectangles).}
\end{center}
\end{figure}

\begin{figure}[t]
\begin{center}
\includegraphics[height=6.0cm,width=6.0cm,angle=270]{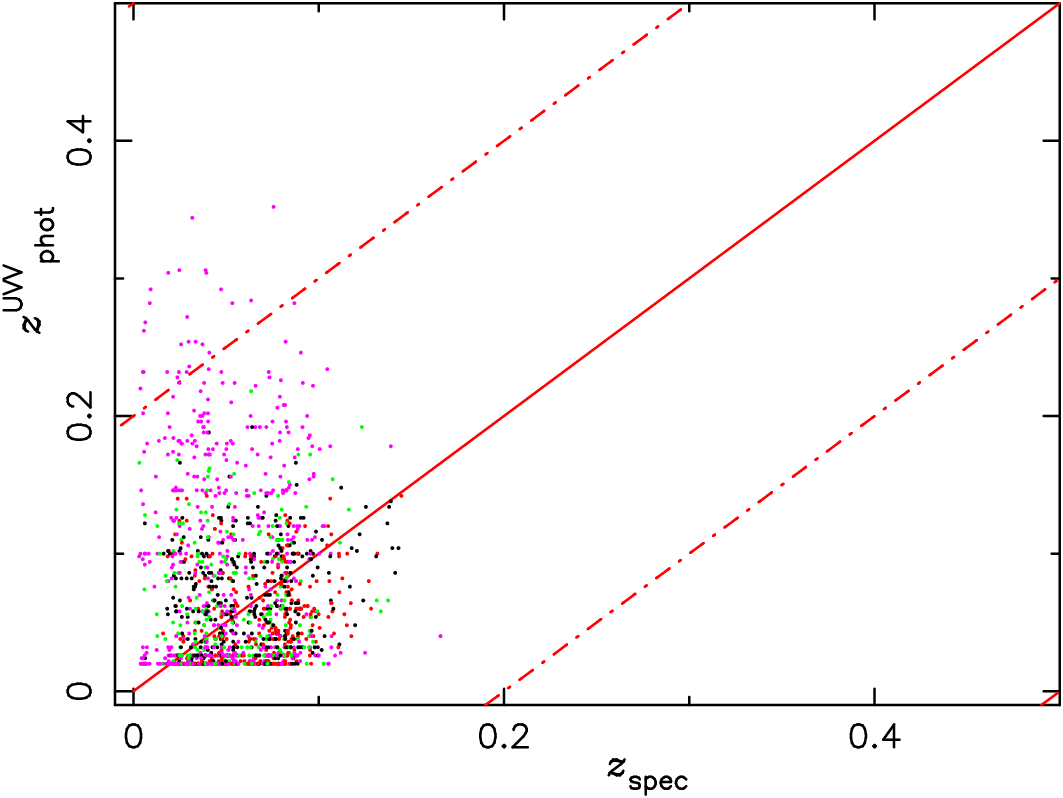}
\caption{Comparison between the SDSS spectroscopic ($z_{\rm spec}$) and photometric ($z^{\rm UVV}_{\rm phot}$) redshifts for 1502 galaxies. Dot-dashed line corresponds to $\delta z$ =0.2. Red, black, green and purple symbols denote galaxies with the $z^{\rm UVV}_{\rm phot}$ probability: $p(\chi^2) \ge 99$, $99 > p(\chi^2) \ge 90$, $90 > p(\chi^2) \ge 68$ and $p(\chi^2) < 68$.}
\end{center}
\end{figure}

\begin{figure}[t]
\begin{center}
\includegraphics[height=6.0cm,width=6.0cm,angle=270]{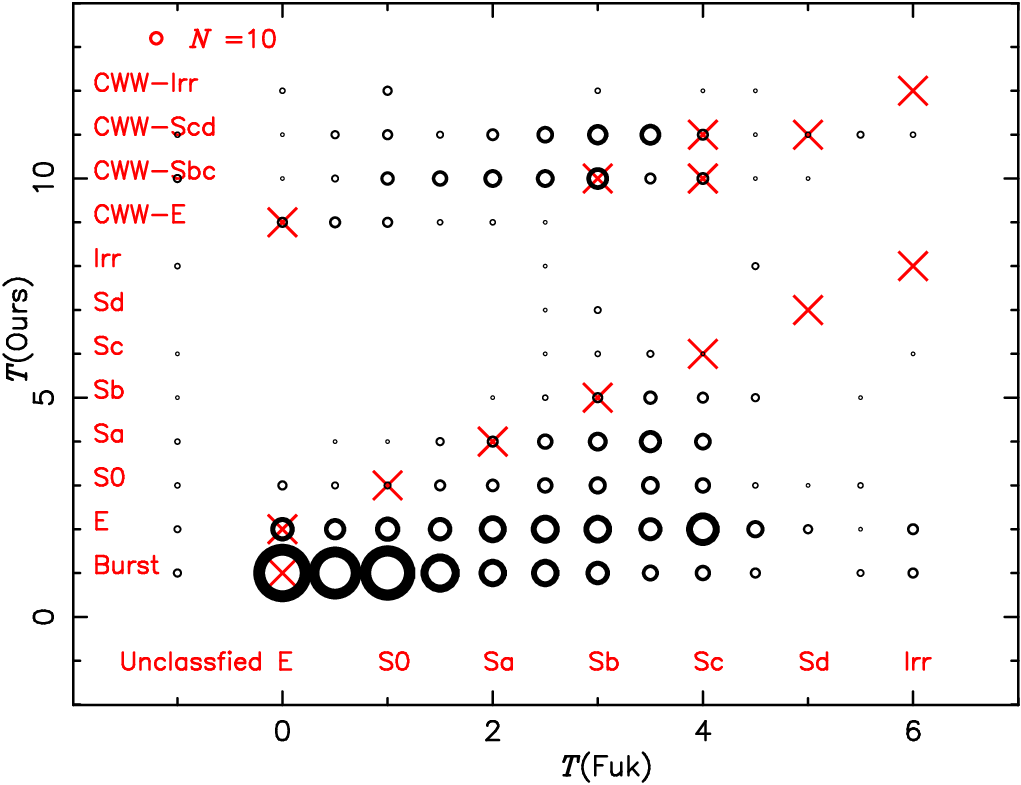}
\caption{Comparison of our $T$ with those of Fukugita et al. for 1502 galaxies. The area of the circle represents the number of galaxies in the grid. Crosses show the correlation in the definition $T$ between ours and Fukugita et al. (2007). {\sl The area of circle located in the upper left corner denotes $N=10$.}}
\end{center}
\end{figure}

\begin{table}[t]
\caption{Morphological index $T$ of Fukugita et al. (2007) and Ours.}
\centering{
\begin{tabular}{lcc}
\hline
galaxy Type  & $T$(Fuk) & $T$(Ours) \\
\hline
Unclassified &   -1     &     ...   \\
 Burst       &  ...     &     1     \\
 E           &   0      &     2     \\
 S0          &   1      &     3     \\
 Sa          &   2      &     4     \\
 Sb          &   3      &     5     \\
 Sc          &   4      &     6     \\
 Sd          &   5      &     7     \\
 Irr         &   6      &     8     \\
 CWW-E       &  ...     &     9     \\
 CWW-Sbc     &  ...     &     10    \\
 CWW-Scd     &  ...     &     11    \\
 CWW-Irr     &  ...     &     12    \\
\hline
\end{tabular}
}
\end{table}

%____________________________________________________________________________
In this part we use this set of SEDs to { determine} redshift and galaxy morphology. In Sections 3.1 and 3.2 we describe the method and the galaxy sample, respectively. { In} Section 3.3 we present the results and the comparisons.

%____________________________________________________________________________
\subsection{Method}\label{ss:metho}
The photo-$z$ is computed through the \textit{Hyperz} code of Bolzonella et al. (2000), which adopts a standard SED fitting method:
\begin{equation}
\chi^2=\sum_{i=1}^{N_{\rm filters}} \Bigl[ {F_{{\rm obs},i}-b \times F_{{\rm temp},i} \over \sigma_i} \Bigr] ^2
\end{equation}
where $F_{{\rm obs},i}${ ,} $F_{{\rm temp},i}$ and $\sigma_i$ are the observed and template fluxes and their uncertainty in filter $i${, respectively,} and b is a constant.
The parameters introduced in the \textit{Hyperz} code include the filter set and the template spectra. In the following computations the template SED library consists of
4 observed mean spectra of local E-, Sbc-, Scd-{ ,} and Irr-type { galaxies} from Coleman, Wu \& Weedmam (1980, hereafter CWW-E, CWW-Sbc, CWW-Scd{ ,} and CWW-Irr)
and 8 { spectral} families built with the HPL07+Yunnan models, and the filter set comprises GALEX $F_{\rm UV}$, $N_{\rm UV}$ and SDSS $ugriz^0$ (for a null airmass from the BC03 package).

Using the BC03 package and the HPL07+Yunnan models, we build 8 different star formation rates, corresponding to the galaxy types from E to Irr, and present the corresponding { spectral} family { for each}. During the construction we suppose that gas does not { get} recycled into new star formation once it { has been} processed into stars and the attenuation by dust is neglected. In Fig.~3 we present the evolution of $F_{\rm UV}$ magnitude of these 8 galaxy types (normalized to 1$M_\odot$), which are built with the Yunnan and HPL07+Yunnan models, respectively.

%____________________________________________________________________________
\subsection{Sample and data}\label{ss:sampl}
Our initial sample of galaxies { is} from the catalogue of Fukugita et al. (2007), which contains 2253 galaxies with Petrosian magnitude $r_p$ brighter than 16\,mag in the north equatorial stripe from the SDSS/DR3.
Removal the objects { mismatched} with {\sl the} SDSS DR7 and GALEX DR4 { produces} a final sample of 1502 galaxies. The matching radius between SDSS/DR7 and GALEX/DR4 is 6.0 arcsec, and the nearest neighbour is taken as a true association if its radius is smaller than the matching radius.

From the catalogue of Fukugita et al. (2007) we get their morphological classification which { is} obtained by visual inspection of images in the $g$ band.
From {\sl the} SDSS/DR7 and GALEX/DR4 we obtain the SDSS spectroscopic redshifts, $F_{\rm UV}$, $N_{\rm UV}$, $u,g,r,i$ and $z$ magnitudes and their errors for these galaxies.

%____________________________________________________________________________
\subsection{Results concerning the {\sl photo-$z$s and} morphologies of galaxies}\label{ss:resul}
Inputting { the} magnitudes { (and their errors)} of { the} 1502 galaxies, the filter set and the template spectra in the \textit{Hyperz} code, we obtain their photo-$z$s and morphologies. In Fig.~4 we give the comparison between the SDSS spectroscopic $z_{\rm spec}$ and the derived photometric redshifts $z^{\rm UVV}_{\rm phot}${ . From} it we see that the $z^{\rm UVV}_{\rm phot}$s fluctuate around $z_{\rm spec}$s{, and the difference between the photometric and spectroscopic redshifts is within 0.2 for most of galaxies}.

In Fig.~5 we give the comparison between our derived morphological index $T$(Ours) with those of Fukugita et al., { 2007,} $T$(Fuk). The corresponding galaxy types of $T$(Ours) and $T$(Fuk) are presented in Table 1. Note in the catalogue of Fukugita et al. (2007) the half-integer classes are allowed for.
From the comparison we see that { the early-type galaxies can be easily and reliably retrieved, while} Sa-Sc galaxies in the catalogue of Fukugita et al. (2007) are { often} classified somewhat earlier, as Burst-E in our classifications. This is caused by that at $z \le 0.2$ no spectral signature can be used in the \textit{Hyperz} code, thus leading to the degeneracy among the fit parameters.

{ On the whole from the comparisons in Figs. 4 and 5 we conclude that the photo-$z$s and the early types of bright galaxies can be reliably obtained, while the late types are often misidentified as early types. Meanwhile, other objects (such as, blue stragglers) from binary interactions also can affect the determination of the photometric redshifts and morphologies of galaxies.}

%____________________________________________________________________________
\section{Summary}\label{s:summa}
In this study we present the SEDs of solar-metallicity BSPs by combining the EPS models of HPL07 and { the} Yunnan group. The set { of} SEDs { is} available on request from the authors. Inclusion of sdBs { makes} the spectra in the ultraviolet passband bluer at late ages.

Using this set of SEDs of BSPs and \textit{Hyperz} code we obtain the photo-$z$s and morphologies of 1502 galaxies. This sample of galaxies is obtained by removing those objects { mismatched} with SDSS/DR7 and GALEX/DR4 from the catalogue of Fukugita et al. (2007).
By comparing the derived photo-$z$s with SDSS spectroscopic redshifts we find that the photo-$z$s fluctuate around the spectroscopic values. By comparing the morphological index with those of Fukugita et al. (2007) we find that the Sa-Sc galaxies in their catalogue are classified earlier in this study. This maybe implies that we should add other feature(s) or structural parameters{, which can be separate early-type galaxies from spirals in advance,} in the galaxy morphology determination.

%____________________________________________________________________________
%____________________________________________________________________________
%% Acknowledgements
%
\acknowledgments
This work was funded by the Chinese Natural Science Foundation
(Grant Nos 10773026, 110673029, 10821061 \& 2007CB15406), { the}
Yunnan Natural Science Foundation (Grant No. 2007A113M) and the Chinese Academy of Sciences (KJCX2-SW-T06). { We thank an anonymous referee for his/her useful suggestions.}

Funding for the SETS and SDSS-II has been provided by the Alfred P. Sloan Foundation, the Participating Institutions, the National Science Foundation, the U.S. Department of Energy, the National Aeronautics and Space Administration, the Japanese Monbukagakusho, the Max Planck Society, and the Higher Education Funding Council for England. The SDSS Web Site is http://www.sdss.org/.

The SDSS is managed by the Astrophysical Research Consortium for the Participating Institutions. The Participating Institutions are the American Museum of Natural History, Astrophysical Institute Potsdam, University of Basel, University of Cambridge, Case Western Reserve University, University of Chicago, Drexel University, Fermilab, the Institute for Advanced Study, the Japan Participation Group, Johns Hopkins University, the Joint Institute for Nuclear Astrophysics, the Kavli Institute for Particle Astrophysics and Cosmology, the Korean Scientist Group, the Chinese Academy of Sciences (LAMOST), Los Alamos National Laboratory, the Max-Planck-Institute for Astronomy (MPIA), the Max-Planck-Institute for Astrophysics (MPA), New Mexico State University, Ohio State University, University of Pittsburgh, University of Portsmouth, Princeton University, the United States Naval Observatory, and the University of Washington.

%This publication makes use of data products from the 2MASS, which is a joint project of the University of Massacgysetts abd tge Infrared Processing and Analyse Center/California Institute of Technology, funded by the National Aeronautics and Space Administration and the National Science Foundation.

The galaxy Evolution Explorer (GALEX) is a NASA Small Explorer. The mission was developed in cooperation with the centre National d'Etudes Spatiales of France and the Korean Ministry of Science and Technology.

%____________________________________________________________________________
%____________________________________________________________________________
%% References

%% Please cite all reference entries in the article text using \cite or
%% equivalent command.

%%%  Using BibTeX  (Name-Year style)
%
% \bibliographystyle{spr-mp-nameyear-cnd}  %% BibTeX style
% \bibliography{<bib data>}                %% BibTeX data

\begin{thebibliography}{}
% \bibitem[\protect\citeauthoryear{<author>}{<year>]{ref:?}
%    <ref. entry>
% \bibitem[\protect\citeauthoryear{<author>}{<year>]{ref:?}
%    <ref. entry>
% \bibitem[\protect\citeauthoryear{<author>}{<year>]{ref:?}
%    <ref. entry>
% \bibitem[\protect\citeauthoryear{<author>}{<year>]{ref:?}
%    <ref. entry>
 \bibitem[\protect\citeauthoryear{Bolzonella, Miralles \& Pell}{Bolzonella et al.}{2000}]{bol00}
 Bolzonella M., Miralles J. M. \& Pell\,o R., 2000, A\&A, 363, 476

 \bibitem[\protect\citeauthoryear{Bruzual \& Charlot}{2003}]{bru03}
 Bruzual G., Charlot S., 2003, MNRAS, 344, 1000 { [BC03]}

 \bibitem[\protect\citeauthoryear{Coleman, Wu \& Weedman}{Coleman et al.}{1980}]{col80}
 Coleman G., Wu C., Weedman D., 1980, ApJS, 43, 393 {[CWW]}

 \bibitem[\protect\citeauthoryear{Fukugita et al.}{2007}]{fuk07}
 Fukugita M., Nakamura O., Okamura S., Yasuda N., Barentine J., Brinkmann J., Gunn J., Harvanek M., et al., 2007, AJ, 134, 579

 \bibitem[\protect\citeauthoryear{Han et al.}{2002}]{han02}
 Han Z., Podsiadlowski Ph., Maxted P. F. L., Marsh T. R., Ivanova N., 2002, MNRAS, 336, 449

 \bibitem[\protect\citeauthoryear{Han et al.}{2003}]{han03}
 Han Z., Podsiadlowski Ph., Maxted P. F. L., Marsh T. R., 2003, MNRAS, 341, 669

 \bibitem[\protect\citeauthoryear{Han, Podsiadlowski \& Lynas-Gray}{2007}]{han07}
 Han Z., Podsiadlowski Ph., Lynas-Gray A. E., 2007, MNRAS, 380, 1098 { [HPL07]}

 \bibitem[\protect\citeauthoryear{Zhang et al.}{2004}]{ref:zha04b}
 Zhang F., Han, Z., Li, L., Hurley J. R., 2004, A\&A, 415, 117

 \bibitem[\protect\citeauthoryear{Zhang et al.}{2005}]{ref:zha05}
 Zhang F., Han, Z., Li, L., Hurley J. R., 2005, MNRAS, 357, 1088
\end{thebibliography}

%% Non-BibTeX  (Name-Year style)
%

\end{document}